# Exact Value Solution to the Equity Premium Puzzle

Atilla Aras[a]

[a] Gazi University, Graduate School of Natural and Applied Sciences

Department of Mathematics

Ankara, Türkiye

E-mail: aaras1974@gmail.com

ORCID ID number: 0000-0002-7727-9797

# Exact Value Solution to the Equity Premium Puzzle

**Abstract**

This article's aim is to provide the solution to the equity premium puzzle without using calibrated values. Calibrated values of subjective time discount factor were used in my prior derived models because 4 variables were determined from 3 different equations. Furthermore, calculated values and risk behavior determination of my prior models were compatible with empirical literature. 4 unknown variables are now calculated from 4 different equations in the new derived model in this article. Subjective time discount factor and coefficient of relative risk aversion are found 0.9581 and 1.0319, respectively from the system of equations which are compatible with empirical studies. Micro and macro studies about CRRA value affirm each other for the first time in the literature. Furthermore, equity and risk-free asset investors are pinned down to be insufficient risk-loving, which can be considered a type of risk-averse behavior. Hence it can be said that calculated values and risk attitude determination align with empirical literature. This shows that derived model is valid and make CCAPM work without calibration.

## 1. Background

Aras (2022, 2024, 2025) derives models to make consumption capital asset pricing model (CCAPM) work. Aras calls these derived models Constant and Variable Sufficiency Factor Models. These models provide a solution to the equity premium puzzle. The CCAPM has not worked because it gives very high coefficient of relative risk aversion (CRRA) under plausible parameters for the historical equity premium of the US financial markets. This reality is coined equity premium puzzle by Mehra and Prescott (1985).

Aras (2022, 2024) calls his derived model Constant Sufficiency Factor Model in his articles. He assumes price to dividend ratio and sufficiency factor of the model (SFOM) constant in the models. He has to use 3 different equations for 4 unknown variables. Hence, he makes use of calibrated value of subjective time discount factor (STDF). He finds CRRA about 1.03 and 1.06 when he assumes STDF 0.99. The risk attitude determination for the model is also compatible with empirical literature.

Aras (2025) derives a new model with time-varying variables and calls this model Variable Sufficiency Factor Model. He uses calibrated STDF values of 0.97, 0.98, and 0.99 in his model. The model assigns the CRRA around 4.40. These values are pinned down to be compatible with empirical literature. Furthermore, risk behavior determination also aligns with empirical studies. Calibrated value using is also relevant for the Variable Sufficiency Factor Model.

We calculate 4 unknown variables from 4 different equations system in this article. Hence, there is no need for calibration. We show the exact value solution to the equity premium puzzle with Constant Sufficiency Factor Model.

The motivation of the article is that there is a research gap in the literature about CCAPM. The research gap can be expressed as follows. There is no study in the literature about CCAPM that gives exact value solution to the equity premium puzzle. The article's contribution to existing literature is that CCAPM now gives exact values of SFOMs, CRRA and STDF under the derived model in this study. Hence, it can be concluded that I reach a solution to the equity premium puzzle without calibrated values.

## 2. Literature Review

Investors compare certain and uncertain utility in the financial investment process. Various risk attitudes emerge at this comparison. SFOM is used when there is future uncertainty and investors have inadequate models for predicting future. It takes the form of coefficient for the uncertain utility. In a way, it defines uncertain utility in terms of certain utility. Investors are then classified as risk-averse and risk-loving when they allocate extra negative and positive utility, respectively for the uncertain wealth.

STDF is less than 1. It shows the impatience of agents in consumption.

CRRA is less than 10 according to empirical studies. Many micro literatures show that it is about 1. According to Levy (2025) it must be close 1. Otherwise, it gives paradoxical results. Elminejad et al. (2025) states that calculated value of 1 is usually used in economics studies and calculated value of 2-7 is frequently used in finance literature.

The fact that standard economic models give an unacceptable CRRA for the equity premium of US financial markets is coined equity premium puzzle by Mehra and Prescott (1985). There has existed no agreed upon solution up till now. Suggested mainstream models can be arranged in three groups. Preference models are habit formation (Abel, 1990) and Epstein-Zin Preferences (Epstein and Zin, 2013). These models alter the calculation of utility. The mainstream paper of behavioral finance models is Myopic Loss Aversion by Benartzi and Thaler (1995). These models modify the assumption of rational agents and look for psychological biases. Key papers of Market Frictions and Rare Disasters are Rare Disasters by Barro (2006) and Idiosyncratic Risk and Borrowing Constraints by Mankiw (1986). These models state that standard models neglect real world obstacles, and their suggested models emphasize possibility low disasters.

Mainstream studies to build the theoretical framework of CCAPM are articles of Lucas (1978) and Breeden (1979). Lucas (1978) stated that asset prices were pinned down by marginal utility of consumption. Additionally, Breeden (1979) formally derived the CCAPM. Hansen and Singleton (1982) derived the GMM to test the CCAPM. Mainstream solutions to fix the CCAPM for equity premium puzzle are studies of Campbell and Cochrane (1999) and Bansal and Yaron (2004). Campbell and Cochrane (1999) introduced the habit formation. Additionally, Bansal and Yaron (2004) introduced the Long-Run Risk Model. The model states that high equity returns are explained by long-run risks in consumption growth.

Efforts to fix CCAPM for the equity premium puzzle have continued over the last 5 years. Andreasen and Jørgensen (2020) suggested new utility kernel to separate IES, RRA and timing attitude. Yan and Wang (2020) suggested heterogeneous consumers to fix the model. Kindrat (2020) investigated how unemployment changed the predictions of CCAPM. Li et al. (2021) modified the CCAPM with bad and good uncertainties. These uncertainties show fat-tailed, skewed negative and positive innovations to the consumption growth. Kim (2021) combined modern portfolio theory with CCAPM to suggest a model for equity premium puzzle. Lee et al. (2021) investigated the relationship between asset returns and consumption growth under the constraint of CCAPM. They based their investigation on maximum entropy method. Zheng et al. (2022) stated that capability of the model in explaining risk was improved by including real economy variables to CCAPM. Aras (2022, 2024, 2025) derived modified CCAPMs that gave CRRA around 1.03 when STDF was assumed 0.99. Price to dividend ratio and SFOMs are assumed constant in his model. The variable version of the model gives CRRA around 4.40 when STDFs are assumed 0.97, 0.98 and 0.99. Snigaroff and Wroblewski (2023) expressed that CCAPM that included earnings and liquidity fitted data better than standard recursive preference models. Aquino et al. (2024) studied discrete-time CCAPM under expectations-based reference-dependent preferences. The model gives asset prices that are align with empirical calculations. Van es (2025) modified the CCAPM by adding banking sector. The writer states that equity premium exists partly because of the banking sector. Zareian Baghdad Abadi et al. (2025) examined the CCAPM by 2 variables. They conclude that surplus consumption ratio is better for short-run predictions and consumption-wealth ratio works for long-run.

We can easily envisage according to the existing literature that efforts to reach a solution to the equity premium puzzle by modified CCAPM or by other models will continue until an agreed solution is possessed.

## 3. Data and Constant Sufficiency Factor Model

### 3.1 Data

I use Mehra and Prescott (1985) data and Table 1 of Mehra (2003). Data subsumes time series data of economic and financial variables for the 1889-1978 period of US Economy.

### 3.2 Sufficiency Factor Model with Constant Price to Dividend Ratio

Sufficiency Factor Model is the first in the literature to state that investors automatically assign extra positive or extra negative utility to uncertain utility through sufficiency factor of the model in comparing certain and uncertain utility in the financial investment process.

Mehra (2003) assumes the price to dividend ratio constant. Aras (2022, 2024) also assumes this ratio constant by presuming SFOMs for equity and risk-free asset investors constant in his Constant Sufficiency Factor Model. He formulates the problem of the typical agent to have the solution for the equity premium puzzle as follows with Constant Sufficiency Factor Model:

$$max_{\{\nabla_{t+1},\ \Phi_{t+1},\ c_t\}} \{v(c_t) + \gamma E_t[\sum_{m=t}^{\infty} \beta^{m+1-t}\, v(c_{m+1})]\}$$

s.t.

$$\Phi_{t+1} f_t + \nabla_{t+1} p_t + c_t \leq \nabla_t y_t + \nabla_t p_t + \Phi_t, \qquad (1)$$

$$c_t \geq 0,\ 0 \leq \nabla_t \leq 1 \text{ for each t.}$$

This problem is solved by dynamic programming. Dynamic programming transforms the problem into two-term periods problem. The typical agent compares certain and uncertain utility each period and reaches a decision in the financial investment process. Then, their risk attitudes are specified according to the new definitions in Aras (2024). Moreover, the dynamic programming and other methods produce the equation system of 2-5. The budget constraint in equation 1 makes use of the principle of available of resources is equal to use of these resources.

We have SFOM for risk-free asset at time t for two reasons:

1-Having the uncertainty of $c_{t+1} = \nabla_{t+1} y_{t+1} + \nabla_{t+1} p_{t+1} + \Phi_{t+1} - \Phi_{t+2} f_{t+1} - \nabla_{t+2} p_{t+1}$ or $c_{t+1} = \nabla_{t+1} y_{t+1} + \nabla_{t+1} p_{t+1} + \Phi_{t+1} R_{f,t+2} - \Phi_{t+2} - \nabla_{t+2} p_{t+1}$ at time t.

2- The fact that risk-free asset investors may trade with the FED (The Federal Reserve System) in the open market during the year produces an uncertainty at time t.

The total economy is in equilibrium

1- when market clearing for equity and risk-free asset investors occurs with $\Phi_{t+1} = 0$ and $\nabla_t = \nabla_{t+1} \ldots = 1$.

2- when x = z occurs at equilibrium.

We need not assume concave utility curve for the risk-aversity. Convex or linear utility curves may also denote risk-aversity for the new definitions. Despite this, I presume concave utility curve for the typical agent because risk-aversity also occurs when the typical agent possesses concave curve in the new definitions. Detailed information about this can be found at Aras (2023).

EXACT VALUE

We have the following equations of Sufficiency Factor Model after including the SFOM to standard CCAPM in Mehra (2003) when the SFOM and price to dividend ratio are presumed constant.

$$\ln E(R_{f,t+1}) = -\ln \beta - \ln \Omega + \tau \mu_x - 0.5\tau^2 \sigma_x^2, \tag{2}$$

See the derivation of equation 2 in Aras (2022) for the proof.

$$\ln E(R_{f,t+1})(1 - \tau \rho \sigma_x \sigma_r) - \ln E(R_{e,t+1}) = -\ln \beta (\tau \rho \sigma_x \sigma_r) + \ln \delta (1 - \tau \rho \sigma_x \sigma_r) - \ln \Omega(1 + \tau \rho \sigma_x \sigma_r), \tag{3}$$

See the derivation of equation 3 in the Appendix for the proof.

$$\ln E(R_{e,t+1}) - \ln E(R_{f,t+1}) = \ln \Omega - \ln \delta + \tau \sigma_x^2, \tag{4}$$

See the derivation of equation 4 in Aras (2022) for the proof.

$$\ln E(R_{e,t+1}) = \ln E(x_{t+1}) - \ln \beta - \ln \delta - (1 - \tau)\mu_x - 0.5(1 - \tau)^2 \sigma_x^2. \tag{5}$$

See the derivation of equation 5 in Aras (2022) for the proof.

Here,

(1) $v(c, \tau) = \frac{c^{1-\tau} - 1}{1-\tau}$;

(2) $\tau$ = coefficient of relative risk aversion;

(3) $c$ is per capita real consumption;

(4) $R_{e,t+1} = \frac{p_{t+1} + y_{t+1}}{p_t}$,

(5) $p_t$ and $y_t$ denote price of the stock and dividend at time t;

(6) $R_{f,t+1} = \frac{1}{f_t}$, where $f_t$ denotes the price of the risk-free asset at time t;

(7) the growth rate of consumption, $x_{t+1} = \frac{c_{t+1}}{c_t}$ ;

(8) the growth rate of dividends, $z_{t+1} = \frac{y_{t+1}}{y_t}$ ;

(9) $(x_t , z_t)$ are jointly lognormally distributed;

(10) E $(R_{e,t+1})$ = mean equity return;

(11) $\mu_x$ = E (ln $x$);

(12) $\sigma_x^2$ = var (ln $x$);

(13) $\beta$ = subjective time discount factor;

(14) $\Omega$= sufficiency factor of the model for risk-free asset investors;

(15) $\delta$= sufficiency factor of the model for equity investors;

(16) $\sigma_r^2$ = var (ln $R_{e,t+1}$);

(17) $\rho$ = corr (lnx, lnr) ;

(18) $(x_t , R_{e,t})$ are jointly lognormally distributed;

(19) $\nabla_t$ = amount of equity;

(20) $\Phi_t$ = amount of risk-free asset;

(21) $\gamma$= sufficiency factor of the model (in the typical agent problem);

(22) E $(R_{f,t+1})$ = mean risk-free rate;

Equations involve SFOMs, CRRA, and STDF. We estimate them from the equation 2-5 of this study.

## 4. Results and Discussion

Sufficiency Factor Model with constant variables is tested with standard tests (i.e., conditional moments are changed with sample means, variances etc.) The solution of the equation system 2-5 is calculated by MATLAB. The results are demonstrated in Table 1 and Table 2.

Calculated CRRA value in this table is compatible with the existing empirical literature. Micro and macro literature about CRRA values, for the first time in finance literature, coincide. The model assigns the CRRA 1.0319. Calculated STDF value also aligns with empirical studies. Moreover, SFOM for risk-free asset and equity investors is more than 1. Hence it can be said that both risk-free asset and equity investors allocate extra positive utility to uncertain wealth. Lastly, both risk-free asset and equity investors are insufficient risk-loving in year 1977, which can be interpreted as a kind of risk-aversity.

None of the suggested models in the literature has stated up till now apart from those of Aras that investors assign extra positive or negative utility to uncertain wealth when they compare certain and uncertain utility in the financial investment process and then are classified as risk-averse or risk-loving according to this allocation. Sufficiency Factor Model is the first to state this allocation.

Both calculated values of CRRA and STDF and risk-attitude determination show that Sufficiency Factor Model with exact value is valid like earlier Sufficiency Factor Models (Aras 2022, 2024, 2025).

**Table 1**

**Calculation Results for Equity Investors**

| STDF | SFOM | CRRA | Certain Utility | Uncertain Utility | Type of investor Year 1977 |
|---|---|---|---|---|---|
| 0.9581 | 1.0013 | 1.0319 | 7.14871804 | 6.27558270 | Insufficient risk- loving |

EXACT VALUE

**Table 2**

**Calculation Results for Risk-free Asset Investors**

| STDF | SFOM | CRRA | Certain Utility | Uncertain Utility | Type of investor Year 1977 |
|---|---|---|---|---|---|
| 0.9581 | 1.0657 | 1.0319 | 7.14871804 | 6.97944955 | Insufficient risk- loving |

## 6. Conclusions

I derive the new model by including the SFOM for risk-free asset and equity to standard CCAPM. I possess 4 different equations and variables under the Constant Sufficiency Factor Model. Hence, I estimate the exact values of SFOMs, STDF, and CRRA. The derived model assigns the CRRA and STDF 1.0319 and 0.9581, respectively. These calculated values align with the empirical literature. Furthermore, micro and macro studies about CRRA values have verified each other for the first time in the literature. Risk attitude determination also confirms that the new derived model is valid. Risk-free asset and equity investors are found insufficient risk-loving, which can be interpreted as a type of risk-aversity. These behavior determinations are also compatible with empirical studies. Hence, it can be stated that CCAPM is now working with exact values under the derived model.

Standard CCAPM with exact value specifying under the derived model gives solution to the equity premium puzzle. Working CCAPM with exact values is very important for economics and finance literature. Policy making with exact value CCAPM will be significant for both real and financial sectors because one can pin down the types of investors easily in financial markets now. Policy making on the connections between real and financial sectors with working CCAPM is also possible now.

I need not possess robust tests or sensitivity analysis for the derived model because I reach the exact value solution for 4 equations with Constant Sufficiency Factor Model. This feature of the Constant Sufficiency Factor Model also strengthens the validity of the model. Hence it can be concluded that all versions of Sufficiency Factor Model provide solution to the equity premium puzzle.

## 7. Statements and Declarations

**Competing Interests:** The author has no relevant financial or non-financial interests to disclose.

**Data Availability Statement:** The data and programs generated (i.e., replication package) for the article are available at https://doi.org/10.7910/DVN/KW5D7S, Harvard Dataverse, V1

**Funding**: No funds, grants, or other support was received.

## Appendix

### Derivation of Equation 3

We have the following from Aras (2025) when the sufficiency factor of the model is constant

$$\Omega E_t(R_{f,t+1}) - \delta E_t(R_{e,t+1}) = \Omega\delta\beta E_t(R_{f,t+1})\, cov_t\left(\frac{u'(c_{t+1})}{u'(c_t)}, R_{e,t+1}\right). \qquad \text{(A.1)}$$

We will use

$$\text{cov}(X^a, Y^b) = \text{E}(X^a)\,\text{E}(Y^b)\,[\exp(ab\rho\sigma_x\sigma_y) - 1] \text{ where } \text{E}(X^a) = \exp(a\mu_x + 0.5a^2\sigma_x^2),\ \text{E}(Y^b) = \exp(b\mu_y + 0.5b^2\sigma_y^2),\ \rho = \text{corr}(\ln x, \ln y) \qquad \text{(A.2)}$$

for $cov_t\left(\frac{u'(c_{t+1})}{u'(c_t)}, R_{e,t+1}\right)$.

Then we have

EXACT VALUE

$$cov_t(\frac{u'(c_{t+1})}{u'(c_t)}, R_{e,t+1}) =$$

$$[\exp(-\tau\mu_x + 0.5\tau^2\sigma_x^2)]\ [\exp(\mu_r + 0.5\sigma_r^2)]\ [\exp(-\tau\rho\sigma_x\sigma_r) - 1], \quad \text{(A.3)}$$

if we use lognormal properties and switch conditional covariance with sample covariance which is align with standard tests.

Substitute A.3 in A.1 to possess

$$\Omega E(R_{f,t+1}) - \delta\ E(R_{e,t+1}) = \Omega\delta\beta E(R_{f,t+1})[\exp(-\tau\mu_x + 0.5\tau^2\sigma_x^2 + \mu_r + 0.5\sigma_r^2 - \tau\rho\sigma_x\sigma_r)]$$

$$-\Omega\delta\beta E(R_{f,t+1})\ [\exp(-\tau\mu_x + 0.5\tau^2\sigma_x^2 + \mu_r + 0.5\sigma_r^2] \quad \text{(A.4)}$$

after using of lognormal properties and switching conditional expectation with sample average which is align with standard tests.

Take ln of both sides to possess

$$\ln\Omega + \ln E(R_{f,t+1}) - \ln\delta - \ln E(R_{e,t+1}) = (\ln\beta + \ln\Omega + \ln\delta + \ln E(R_{f,t+1})\ (-\tau\mu_x + 0.5\tau^2\sigma_x^2 + \mu_r + 0.5\sigma_r^2 - \tau\rho\sigma_x\sigma_r) - (\ln\Omega + \ln E(R_{f,t+1}) + \ln\delta + \ln\beta)\ (-\tau\mu_x + 0.5\tau^2\sigma_x^2 + \mu_r + 0.5\sigma_r^2).$$

(A.5)

Some algebraic operations after A.5 result in

$$\ln E(R_{f,t+1})\ (1 - \tau\rho\sigma_x\sigma_r) - \ln E(R_{e,t+1}) = -\ln\beta\ (\tau\rho\sigma_x\sigma_r) + \ln\delta\ (1 - \tau\rho\sigma_x\sigma_r) - \ln\Omega(1 + \tau\rho\sigma_x\sigma_r)\ .$$

(A.6)